\documentclass[showpacs,prb]{revtex4}

\usepackage{amsmath}
\usepackage{amssymb}

\begin{document}

\title{Interacting potential between spinons in the compact
$QED_3$ description of the Heisenberg model}
\author{Raoul Dillenschneider}
\email[E-mail address : ]{rdillen@skku.edu}
\affiliation{
Department of Physics and Institute for Basic Science Research,
Sungkyunkwan University, Suwon 440-746, Korea
}
\author{Jean Richert}
\email[E-mail address : ]{richert@lpt1.u-strasbg.fr}
\affiliation{
Laboratoire de Physique Th\'eorique, Universit\'e Louis Pasteur,
67084 Strasbourg Cedex, France
} 

\date{\today}

\begin{abstract}
We implement a Chern-Simons (CS) contribution into the compact $QED_3$
description of the antiferromagnetic Heisenberg model in two dimensions
at zero temperature. The CS term allows for the conservation of the 
$SU(2)$ symmetry of the quantum spin system and fixes the flux through a
plaquette to be a multiple of $\pi$ as was shown by Marston. 
We work out the string tension of the confining potential which acts 
between the spinons and show that the CS term induces a screening 
effect on the magnetic field only. The confining potential between spinons
is not affected by the CS flux.
The strict site-occupation by a single spin $1/2$  is enforced by the 
introduction of an imaginary chemical potential constraint.
\end{abstract}


\maketitle

\section{Introduction}

Quantum phase transitions of matter near zero temperature have attracted
much interests in the recent past. A possible mechanism for high-$T_c$
superconductivity may be a transition between an antiferromagnetic N\'eel 
phase and a valence-bond-solid (VBS) phase, see f.i. ref.
\cite{LeeNagaosaWen-04}. Frustrated Heisenberg interactions can be mapped into 
a non-linear sigma model 
from which it is shown that topological defects play an important role in 
the spinon deconfinement through the phase transition from a N\'eel phase to
a VBS phase \cite{Senthil}.
We shall introduce below gauge theories which cannot predict such kind of
phase transitions in non-frustrated Heisenberg models.

At low energy non-frustrated Heisenberg Hamiltonian can be reduced
to Dirac actions. Indeed, a gauge field formulation of the antiferromagnetic 
Heisenberg model in $d = 2$ dimensions leads to a quantum electrodynamic
$QED_3$ action for spinons \cite{PRB144404}. 
It was shown through a renormalization group study of compact 
$(2+1)$-dimension\-nal Maxwell electrodynamics coupled to fermion field with 
$SU(N)$ symmetry that the fermions cannot deconfine when $N$ is lower than 
$20$ \cite{NogueiraKleinert_condmat07}.
This is of peculiar interest for the $QED_3$ description of the
non-frustrated Heisenberg model. Indeed, in the latter case the number
of replica is $N=2$ which implies that the spinons will not deconfine
and the Heisenberg model will not present a paramagnetic phase (i.e. no VBS
phase). We shall provide here arguments which agree with 
\cite{NogueiraKleinert_condmat07} and are based on the introduction
of a Chern-Simons term into to compact $QED$ description of the non-frustrated
Heisenberg models.

We consider the $\pi$-flux state approach introduced by Affleck and 
Marston \cite{aff1,aff2}. In this description it was shown  
that the flux through a plaquette formed by four spin sites must 
be equal to multiples of $\pi$ in order to satisfy the projection properties 
of the loop operator \cite{Marston}. The flux can be strictly fixed to $k \pi$ 
where $k$ is an integer by means of a Chern-Simons (CS) term. We introduce 
such a term here in order to fix the flux and assure the 
conservation of the $SU(2)$ symmetry of the quantum spin system.

It is well known that in compact Maxwell theory Dirac magnetic monopoles
(instantons) in (2+1) dimensions lead to confinement of test particles
\cite{Polyakov}. The question now arises about the effects produced by the
introduction of a Chern-Simons term in the compact $\pi$-flux description of 
the Heisenberg interaction. We shall review well known results which lead 
to the conclusion that the flux through a plaquette controled by the CS term,
screens only the magnetic field between spinons but does not affect the 
confining potential.

In the present approach the spin site-occupation is strictly fixed to one 
through the introduction of an imaginary chemical potential \cite{Popov-88} 
avoiding the introduction of a Lagrange multiplier term \cite{PRB024409}.

The outline of the paper is as follows. In section \ref{Section1} we 
recall the main steps of the $QED_3$ formulation of the two-dimensional
antiferromagnetic Heisenberg model. A justification for the implementation
of the CS term is given and the modification induced by the presence 
of instantons is discussed. Section \ref{Section2} deals with the derivation 
of the instanton action.  In section \ref{Section3} the string tension
of the potential between spinons is worked out.

\section{Flux constraint in the presence of topological defects 
\label{Section1}}

Heisenberg quantum spin Hamiltonians of the type

\begin{equation}
H = -\frac {1}{2}\sum_{i,j} J_{ij} \vec S_{i} \vec S_{j}
\label{eq1}
\end{equation}

\noindent
with antiferromagnetic coupling $\{J_{ij}\} < 0$ can be mapped onto Fock space 
by means of the transformation 
$S^{+}_{i} = f^{\dagger}_{i, \uparrow} f_{i, \downarrow}$,
$S^{-}_{i} = f^{\dagger}_{i, \downarrow} f_{i, \uparrow}$ and
$S^{z}_{i} = \frac{1}{2} (f^{\dagger}_{i,\uparrow} f_{i,\uparrow} - 
f^{\dagger}_{i,\downarrow} f_{i,\downarrow})$ 
where $\{f^\dagger_{i, \sigma}, f_{i, \sigma}\}$ are anticommuting 
fermion operators which create and annihilate spinons with $\sigma=\pm 1/2$.
The projection onto Fock space is exact when the number of fermions per 
lattice site verifies 
$\underset{\sigma=\pm 1/2}{\sum} f^\dagger_{i,\sigma} f_{i,\sigma}=1$.
This is enforced here by using the Popov and Fedotov procedure
\cite{Popov-88,PRB024409} which introduces the imaginary chemical
potential $\mu = i \pi / 2 \beta$ at temperature $\beta^{-1}$.

The Hamiltonian given by equation \eqref{eq1} is invariant under $SU(2)$
symmetry and also under the $U(1)$ gauge transformation 

\begin{equation}
f_{i,\sigma} \rightarrow f_{i,\sigma} e^{i g \theta_i}
\label{eq2}
\end{equation}

\noindent
In 2$d$ space the Heisenberg interaction can be written in terms of a
$\pi$-flux mean-field Hamiltonian for which the mean-field flux $\phi_{mf}$ 
through a square plaquette of four spin sites is given by

\begin{equation*}
\phi_{mf} = g \underset{<ij> \in \Box}{\sum}
\left(\theta_i - \theta_j \right) = \pi m
\end{equation*}

\noindent
where $\theta_i$ is the gauge phase appearing in the gauge transformation
\eqref{eq2} and $m$ is an integer.
The $\pi$-flux mean-field ansatz keeps the Hamiltonian \eqref{eq1} invariant
under $SU(2)$ symmetry transformations. The dispersion relation of the 
$\pi$-flux mean-field Hamiltonian shows two independent nod\-al points. 
Near these nodal points the dispersion relation is linear with respect to 
the momentum vector \cite{PRB144404}.

In the neighbourhood of the nodal points and at low energy the Hamiltonian 
\eqref{eq1} can be rewritten in terms of a four-component Dirac spinon 
action in the continuum limit \cite{GhaemiSenthil-05,Morinari-05,PRB224443}.
This action describes a spin liquid in (2+1) dimensions which includes the 
phase fluctuations $\delta \phi$ around the $\pi$-flux mean field phase 
$\phi_{mf}$. It has been derived in \cite{PRB224443} and reads

\begin{eqnarray}
S_{E} = \int_0^\beta d\tau \int d^2 r
&\Bigg\{&
- \frac{1}{2} a_\mu \left[ \left(\Box \delta^{\mu \nu}
+ (1 - \lambda) \partial^\mu \partial^\nu \right) \right] a_\nu
\notag \\
&+& \underset{\sigma}{\sum}
\bar{\psi}_{\vec{r} \sigma} \left[
\gamma_\mu \left( \partial_\mu - i g a_\mu \right)
\right] \psi_{\vec{r} \sigma}
\Bigg\}
\label{SpinonGaugeAction}
\end{eqnarray}

\noindent
In the following we consider the zero temperature limit 
$\beta \rightarrow \infty$.
Here $a_\mu = \partial_\mu \theta$ is a gauge field generated from the 
$U(1)$ symmetry invariance of $S_{E}$ when $\psi \rightarrow  e^{i g \theta} 
\psi$. The bi-spinor Dirac spinon field
\noindent
$\psi$ 
is defined by
\begin{eqnarray*}
\psi_{\vec{k} \sigma} = \left(
\begin{array}{c}
f_{1 a, \vec{k} \sigma} \\
f_{1 b, \vec{k} \sigma} \\
f_{2 a \vec{k} \sigma} \\
f_{2 b \vec{k} \sigma}
\end{array}
\right)
\end{eqnarray*}

\noindent
where $f^\dagger_{1,\vec{k},\sigma}$ and $f_{1,\vec{k},\sigma}$ 
($f^\dagger_{2,\vec{k},\sigma}$ and $f_{2,\vec{k},\sigma}$) 
are fermion creation and annihilation operators which act near the
nodal points
$(\frac{\pi}{2},\frac{\pi}{2})$ ($(-\frac{\pi}{2},\frac{\pi}{2})$) 
of the momentum $\vec{k}$. Indices $a$ and $b$ characterize 
the rotated operators

\begin{eqnarray*}
\begin{cases}
f_{a,\vec{k},\sigma} = \frac{1}{\sqrt{2}}
\left(f_{\vec{k},\sigma} + f_{\vec{k}+\vec{\pi},\sigma}\right) \\
f_{b,\vec{k},\sigma} = \frac{1}{\sqrt{2}}
\left(f_{\vec{k},\sigma} - f_{\vec{k}+\vec{\pi},\sigma}\right)  
\end{cases}
\end{eqnarray*}

\noindent
The constant $g$ in 
\eqref{SpinonGaugeAction} is the coupling strength between 
$a_\mu$ and $\psi$. The first term corresponds to the 
``Maxwell'' term $-\frac{1}{4} \mathcal{F}_{\mu \nu} \mathcal{F}^{\mu \nu}$  
where $\mathcal{F}^{\mu \nu} = \partial^\mu a_\nu - \partial^\nu a_\mu$,
$\lambda$ is the parameter of the Faddeev-Popov gauge fixing term
$-\lambda \left(\partial^\mu a_\mu \right)^2$ \cite{Itzykson},   
$\delta^{\mu \nu}$ the Kronecker $\delta$, 
$\Box = \partial_\tau^2 + \vec{\nabla}^2$ the Laplacian in Euclidean 
space-time. This form of the action originates from a shift of the imaginary 
time derivation $\partial_\tau \rightarrow \partial_\tau + \mu$ where $\mu$
is the imaginary chemical potential introduced above. It leads to a 
new definition of the Matsubara frequencies of the fermion fields 
\cite{Popov-88} $\psi$ which then read $\widetilde{\omega}_{F,n} = 
\omega_{F,n} - \mu/i = \frac{2 \pi}{\beta} (n + 1/4)$.

Fluctuations of the flux around the $\pi$-flux mean-field are constrained by 
means of symmetry considerations on the loop operator 
$\Pi = f^{\dagger}_{\vec{i}} f_{\vec{i}+\vec{e}_x}
f^{\dagger}_{\vec{i}+\vec{e}_x} f_{\vec{i}+\vec{e}_x+\vec{e}_y}
f^{\dagger}_{\vec{i}+\vec{e}_x+\vec{e}_y }
f^{\dagger}_{\vec{i}+\vec{e}_y}$ $f_{\vec{i}}$.
As shown by Marston \cite{Marston}, only gauge configurations of the flux 
states belonging to $Z_2$ symmetry ($\pm \pi$) are allowed. Hence the
flux through a four-site plaquette is restricted to 
$\phi_{\square} = \phi^{mf} + \delta \phi 
=\left\{ 0,\pm \pi \right\} (\text{mod } 2\pi)$
. This was derived in the following way \cite{Marston}.
The loop operator verifies $\Pi^3=\Pi$.
Defining two quantum states $|u>=\Pi^2 |\varphi>$ and $|v>=(1-\Pi^2)|\varphi>$
where $|\varphi> = |u> + |v>$ is a general quantum state it is easy to
see that $<v|\Pi|v> = 0$ and $\Pi^2 |u> = |u>$. From the last equality one
deduces that $|u>$ can be decomposed into the eigenstates of $\Pi$ with
eigenvalues $\pm 1$. The loop operator can also be rewritten as
$\Pi = |\Pi| e^{i \phi_{\square}}$ where $\phi_{\square}$ is the total flux 
through the plaquette. 
In order to guarantee the properties of $\Pi$ the
total flux through the plaquette has to verify ${\phi_{\square}} = \pi k$ 
where $k$ is an integer. Other values are thus ``forbidden'' gauge 
configurations.

In order to remove these configurations ($\phi_{\square} \neq \pm \pi$) 
in the case of the Heisenberg antiferromagnet a CS term is introduced
in the $QED_3$ action in order to fix the total flux through the
plaquette. This leads to the  Maxwell-Chern-Simons (MCS) action in Euclidean 
space

\begin{eqnarray}
S_{E} &=& \int_0^\beta d\tau \int d^2 r
\Bigg\{
- \frac{1}{2} a_\mu \Big[ \left(\Box \delta^{\mu \nu}
+ (1 - \lambda) \partial^\mu \partial^\nu \right) 
\notag \\
&&
+ i \kappa \varepsilon^{\mu \rho \nu} \partial_\rho  \Big] a_\nu
+ \underset{\sigma}{\sum}
\bar{\psi}_{\vec{r} \sigma} \left[
\gamma_\mu \left( \partial_\mu - i g a_\mu \right)
\right] \psi_{\vec{r} \sigma}
\Bigg\}
\notag \\
\label{MCSAction}
\end{eqnarray}

\noindent
where $g$ couples the spinon field to the gauge field and $\kappa$ is the CS 
coefficient.
Under normal conditions the CS contribution breaks parity and time-reversal 
invariance. However the CS coefficient $\kappa$ can be chosen in such a way 
that the variation of the CS action under a gauge transformation can be an 
integer multiple of $2\pi$.
Indeed under a gauge transformation $a_\mu \rightarrow a_\mu 
+ \partial_\mu \Lambda$ the variation of the CS action can be rewritten
$\delta S_{CS} = \kappa \int d\Sigma_\mu \left(\Lambda 
\bar{\mathcal{F}}_\mu\right)$ where $\bar{\mathcal{F}}_\mu \equiv \frac{1}{2}
\epsilon_\mu^{\phantom{\mu} \alpha \lambda} \mathcal{F}_{\alpha \lambda}$.
Specializing to the gauge transformation $\Lambda = (2\pi n / \beta)\tau$
where $n$ is an integer and only different from zero inside a plaquette
the variation of the CS action reads $\delta S_{CS} 
= \kappa 2\pi n \int d^2 r \bar{\mathcal{F}}_0$.
The integration is simply equal to the flux passing through a plaquette and
the variation of the CS action is equal to 
$\kappa 2 \pi n \phi_{\square}$. Since the flux must be a multiple of $\pi$
the CS coefficient $\kappa$ can always be chosen such that $\delta S_{CS}
=2\pi m$ where $m$ is an integer \cite{Marston}.
Under this condition the variation of the CS action 
no longer contributes to the path integral and the effects of 
P and T symmetry breaking are avoided \cite{Marston}.
The magnetic field $\mathcal{B}$ through a plaquette is related to the flux 
constraint $\phi_{\square} = \pi (\text{mod }2\pi)$ and can be fixed through 
the CS action with such a specific coefficient $\kappa$ 
\cite{Marston,PRB144404}.
Moreover states of the spin system for which the flux through plaquettes 
is a multiple of $\pi$ are all equivalent and connected through gauge
transformations. The variation of the CS action under such gauge 
transformations does not contribute to the path integral as mentioned before.

Instanton generation from the compactness of the gauge field connects these 
different spin states with fluxoids equal to $2\pi$.
In the compact $QED_3$ description of the $\pi$-flux mean field action 
the symmetry ($\Pi^3=\Pi$) of the loop operator remains unbroken. The flux 
through a plaquette $\phi_\square$ has to be fixed to multiples of $\pi$ even 
in presence of instantons.
The instantons introduce a flux through the plaquette equal to 
$\phi_{inst} = 2\pi q$ where 
the integer $q$ is the total \emph{winding} charge of the instantons in the 
plaquette. The flux through a plaquette is $\phi_\square
= \phi^0 + \phi_{inst}$ where $\phi^0$ is the flux without instantons. It is 
therefore clear that $\phi^0$ has to be fixed to multiples of $\pi$ to ensure
the symmetry of $\Pi$. Hence the Chern-Simons term is introduced to 
control the fluctuations of $\phi^0$ but it does not affect the
fluctuations of the instanton density. 

The compact Maxwell-Chern-Simons (MCS) action for the Heisenberg model then 
reads

\begin{eqnarray}
&&S_{E}^{compact} = 
\notag \\
&&
\int_0^\beta d\tau \int d^2 r
\Bigg\{
- \frac{1}{2} a_\mu \left[ \left(\Box \delta^{\mu \nu}
+ (1 - \lambda) \partial^\mu \partial^\nu \right) \right] a_\nu
\notag \\
&&+ \underset{\sigma}{\sum}
\bar{\psi}_{\vec{r} \sigma} \left[
\gamma_\mu \left( \partial_\mu - i g a_\mu \right)
\right] \psi_{\vec{r} \sigma}
\Bigg\}_{\mathcal{F}_{\mu \nu} \rightarrow \widetilde{\mathcal{F}}_{\mu \nu} }
\notag \\
&&+
\int_0^\beta d\tau \int d^2 r
\frac{1}{2} a_\mu \Big[ -
i \kappa \varepsilon^{\mu \rho \nu} \partial_\rho  \Big] a_\nu
\label{compactMCSAction}
\end{eqnarray}

\noindent
where the compact version of the Maxwell and spinon action is generated 
through the transformation

\begin{equation*}
\mathcal{F}_{\mu \nu} \rightarrow \widetilde{\mathcal{F}}_{\mu \nu} 
= \mathcal{F}_{\mu \nu} - 2\pi n_{x,\mu \nu}
\end{equation*}

\noindent
where $\mathcal{F}$ is the electromagnetic tensor defined above in the absence 
of instantons. In this transformation,
$n_{x,\mu \nu} = \varepsilon_{\mu \nu \gamma} \partial_\gamma \varphi_x$
where $\varphi_x$ is the scalar potential generated by the instanton charge 
$q_x$ through the Poisson equation $\Delta_{x,x^{'}} \varphi_{x^{'}}$ $ = q_x$ 
where $q_x$ is an integer \cite{Polyakov}.

\section{Instanton action with flux-controlled spinon field
\label{Section2}}

Integrating out the matter field $\psi$ the MCS action
\eqref{MCSAction} leads to the definition of the gauge field propagator  
at zero temperature \cite{PRB144404,KleinertNogueira}

\begin{eqnarray}
\Delta_{E,\mu ,\nu} &=& 
\frac{1}{k^2 \varepsilon_{\kappa}(k)}
\Bigg(
\delta_{\mu \nu} - \frac{k_\mu k_\nu}{k^2} 
\notag \\
&&
- \frac{\kappa}{(k^2 + \Pi(k))}
\varepsilon_{\mu \nu \rho} k_\rho 
\Bigg) 
+ \frac{k_\mu k_\nu}{\lambda (k^2)^2}
\notag \\
\label{Propagator}
\end{eqnarray}

\noindent
where $\varepsilon_{\kappa}(k) = 1 + \frac{\Pi(k)}{k^2} 
+ \frac{\kappa^2}{k^2 + \Pi(k)}$ is the dielectric function induced by the 
matter field and flux through plaquettes, $\kappa$ is the CS coefficient
as defined above 
and $\lambda$ the Faddeev-Popov gauge fixing parameter. In this gauge field
propagator $\Pi(k) = \alpha k$ is the polarization contribution at the one-loop
approximation and $\alpha=2 g^2$ the coupling constant between the 
(pseudo)-electromagnetic field and the spinon field considered here as the 
fermionic matter field.

Instantons appear only in the Maxwell and spinon terms when $\mathcal{F}$ 
goes over to $\widetilde{\mathcal{F}}$

\begin{eqnarray*}
&&\int \frac{d^3k}{(2\pi)^3} \varepsilon_{\kappa=0}(k). \frac{1}{4}
\widetilde{\mathcal{F}}_{\mu \nu}(k)^2 
\notag \\
&&= \int \frac{d^3k}{(2\pi)^3} \varepsilon_{\kappa=0}(k). \frac{1}{4}
\left(\mathcal{F}_{\mu \nu}(k) - 2\pi n_{k,\mu \nu} \right)^2 
\end{eqnarray*}

\noindent
which leads to the partition function of the gauge field $a_\mu$ 

\begin{eqnarray*}
\mathcal{Z} &=& \mathcal{Z}_{(0)} \times \mathcal{Z}_{inst}
\end{eqnarray*}

\noindent
where $\mathcal{Z}_{(0)}$ and $\mathcal{Z}_{inst}$ are respectively the
bare electromagnetic and the instanton contribution  to the partition 
function. One obtains 

\begin{eqnarray*}
\mathcal{Z}_{(0)} &=& \int \mathcal{D}a_\mu e^{-\frac{1}{2} 
\int \frac{d^3k}{(2\pi)^3} a_\mu \Delta_{\mu \nu}^{-1} a_\nu }
\end{eqnarray*}

\noindent
The topological defects created by instantons through the compactification
lead to $\mathcal{Z}_{inst}$ given by

\begin{eqnarray}
\mathcal{Z}_{inst} &=& \underset{\left\{q_x\right\}}{\sum}
e^{-\int \frac{d^3k}{(2\pi)^3} 4\pi^2 \varphi_{-k}
\left(k^2 \varepsilon_{\kappa=0}(k)\right) \varphi_{k} }
\label{eq6}
\end{eqnarray}
 
\noindent
where $\varphi_k$ is the Fourier transform of the scalar potential 
$\varphi_{x}$ 
defined above and generated by the integer \emph{winding}
charges $q_x$ over which the sum is performed in equation \eqref{eq6}.
The scalar potential $\varphi_k$ is related to the instanton density 
$\rho_{inst}(x) = \underset{x_a}{\sum} q_a \delta(x-x_a)$ by the Poisson
formula $\varphi_k = \frac{\rho_{inst}(k)}{k^2 \varepsilon_{\kappa=0}(k)}$
where the dielectric function $\varepsilon_{\kappa}(k)$ stems from
the gauge field propagator \eqref{Propagator}.

The partition function \eqref{eq6} can be put in a functional integral form 
\cite{Polyakov}. Performing a Hubbard-Stratonovich (HS) transformation on 
equation \eqref{eq6} with respect to the instanton charges $q_a$ leads to 

\begin{eqnarray}
\mathcal{Z}_{inst} &=& 
\int \mathcal{D} \chi \left(e^{
- \int \frac{d^3k}{(2\pi)^3} 
\chi(-k) \frac{k^2\varepsilon_{\kappa=0}(k)}{4\pi^2} \chi(k)} 
\right)
\notag \\
&&\times \underset{N}{\sum} \underset{\left\{q_a \right\}}{\sum}
\frac{\xi^{N}}{N!} \int \overset{N}{\underset{j=1}{\prod}} dx_j
e^{i \underset{\left\{x_a \right\}}{\sum} q_a \chi(x_a) }
\end{eqnarray}

\noindent
where $\xi$ is the instanton fugacity which is related to the dielectric 
function through $\ln \xi = - \frac{1}{4\pi} \int \frac{d^3k}{(2\pi)^3} 
\frac{1}{\varepsilon_\kappa=0(k)k^2}$.
The auxiliary field $\chi$ is generated by the HS transformation.
Following refs. \cite{Polyakov,KleinertNogueira,NazarioSantiago} we assume 
that $q_a = \pm 1$ are the only relevant instanton charges. Then  

\begin{eqnarray}
\mathcal{Z}_{inst} &=& \int \mathcal{D} \chi e^{
- \frac{1}{(2\pi)^2} \int \frac{d^3k}{(2\pi)^3} 
\left( \chi(-k) k^2\varepsilon_{\kappa=0}(k) \chi(k) \right)}
\notag \\
&&
\times e^{\frac{M^2}{(2\pi)^2} \int d^3x \cos \chi(x) }
\label{eq8}
\end{eqnarray}

\noindent
In this last equation $M^2 = (2\pi)^2 \xi$ induces a confining potential 
between two test particles \cite{Polyakov}. At this point it is interesting 
to make a comparison of \eqref{eq8} with the classical instanton action given 
by Polyakov.
In our case the matter field leads to the 
appearance of a dielectric function in the instanton partition function. This 
dielectric function induces modifications on the string tension between test 
particles as it will be shown in the next section. It affects also the dual
field $H_\mu = \epsilon_{\mu \nu \rho} \mathcal{F}_{\nu \rho}$.

When instantons are present in the system the $H$-field is given by two
terms, the bare electromagnetic field contribution 

\begin{eqnarray*}
H_\mu^{(0)}(k) = \epsilon_{\mu \nu \rho} \mathcal{F}_{\nu \rho}(k) = 
\epsilon _{\mu \beta \gamma} i k_\beta a_\gamma(k)
\end{eqnarray*}

\noindent
and $H_\mu^{inst}$ which stems from the \emph{magnetic} field created by 
instantons 

\begin{eqnarray*}
H_\mu^{inst} = \frac{2\pi i k_\mu \rho_{inst}(k)}
{k^2 \varepsilon_{\kappa=0}(k)}
\end{eqnarray*}

\noindent
The introduction of a matter field as well as a flux through plaquette
controlled by a Chern-Simons coefficient $\kappa$ induces a screening
of the $H$-field as can be seen on the correlation function

\begin{eqnarray*}
&&<H_\mu(-k)H_\nu(k)> =
\notag \\
&&
\frac{1}{\varepsilon_{\kappa}(k)} 
\Bigg(
\delta_{\mu \nu} 
- \frac{k_\mu k_\nu}{k^2}
+ \frac{\kappa}{\left(k^2 + \Pi(k)\right) } \varepsilon_{\mu \nu \rho} k_\rho
\Bigg)
 \notag \\
&&
+ \frac{M^2}
{\varepsilon_{\kappa=0}(k)\left(M^2+k^2\varepsilon_{\kappa=0}(k) \right)}
\frac{k_\mu k_\nu}{k^2}
\label{Correlation}
\end{eqnarray*}

\noindent
where $H_\mu(k) = H_\mu^{(0)}(k) + H_\mu^{inst}(k)$.
In the absence of topological defects $M=0$ it is easy to see that the 
magnetic field is screened with a characteristic length $1/\kappa$ in agreement
with \cite{Diamantini}. In the case $M \neq 0$ the photons are massive 
\cite{Polyakov} and the photon mass $M$ is not affected by the Chern-Simons 
term. From this result we anticipate that the string tension 
between spinons will not be affected by the Chern-Simons term (i.e. the flux 
tied to each spinon and proportionnal to $1/\kappa$).

\section{String tension between two test particles \label{Section3}}

The effective potential between two test particles can be obtained from the
Wilson loop \cite{Wilson}. Given a loop contour $C$, the Wilson loop is a 
gauge invariant $W(C) = < e^{- \oint_C dx_\mu a_\mu(x)} >$ and
leads to the potential $V(R) = - \underset{T \rightarrow \infty}{\lim}$
$\frac{1}{T} \ln W(C)$ \cite{Nagaosav1} where $R$ and $T$ are the lengths 
of the loop $C$ in the $xy$ plane. If the potential is not confining the 
logarithm of the Wilson loop is proportional to $T+R$, this is the so called 
perimeter law, and if the potential is confining it leads to the area law 
$\ln W(C) \propto RT$.   

We shall now show that when a matter field is present the 
Wilson loop follows the area law but the string tension is reduced by 
screening effects.

The Wilson loop operator can be rewritten $W(C) = < e^{i\int H_\mu dS_\mu}>
= <e^{i\int H_\mu^0 dS_\mu}>_{\mathcal{Z}_{(0)}} 
\times <e^{i\int H_\mu^{inst} dS_\mu}>_{\mathcal{Z}_{inst}}$ where the
$H$-field has been separated into the bare gauge field and the instanton 
$H$-field contributions. The average over the bare gauge field
leads to the screened \cite{PRB144404,Luscher} Coulomb interaction and will 
be disregarded here.
The second average leads to the instanton confining potential which is of
interest here. The Wilson loop with respect to the instanton action reads

\begin{eqnarray}
W_{inst}(C) &=& < e^{- \oint_C dx_\mu a_\mu(x)} >_{inst} 
\notag \\
&=& <e^{- \int_{x \in C} dS_\mu H^{inst}_\mu(x)} >_{inst}
\notag \\
&=& \int \mathcal{D} \chi e^{
- \frac{1}{(2\pi)^2} \int \frac{d^3k}{(2\pi)^3} 
\left( \left[\chi_{-k}-\eta_{-k}\right] k^2\varepsilon_{\kappa=0}(k) 
\left[\chi_k-\eta_k\right] \right) }
\notag \\
&& \times
e^{\frac{M^2}{(2\pi)^2} \int d^3x \cos \chi(x) }
\label{eq9}
\end{eqnarray}

\noindent
where $<\dots>_{inst}$ stands as an average induced by the instanton partition 
function $\mathcal{Z}_{inst}$. In equation \eqref{eq9} 
$\eta(-k) = \int dS_x \frac{2\pi i k_\mu}{k^2 \varepsilon_{\kappa=0}(k)}
e^{i k.x}$ and $H^{inst}_\mu(k)$ is given in Section \ref{Section2}.

The Wilson loop can be approximated by the
classical solution $\chi_{cl}$ obtained by a saddle-point method on the 
functional integral \eqref{eq9} and in the limit $R,T \rightarrow \infty$
one gets the classical solution

\begin{eqnarray}
\chi_{cl}(k)
= \frac{- 2\pi i k_z e^{- i k_z z} .(2\pi)^2 \delta(k_x)\delta(k_y)
}{\left( k^2 \varepsilon_{\kappa=0}(k) + M^2 \right) }
\label{eq10}
\end{eqnarray}

\noindent
Here we assumed that $\chi_{cl}$ is sufficiently small so that  
$\cos \chi_{cl} \propto 1 - \frac{1}{2} \chi_{cl}^2$ leading to equation
\eqref{eq11}. The introduction of $\chi_{cl}(k)$ into \eqref{eq9} leads to

\begin{eqnarray}
W_{inst}(C) =
e^{
- g^2 R T \left(- \partial_z^2 \right)
\left(
F\left[ \frac{1}{k^2 \varepsilon_{\kappa=0}(k)} \right]_{z}
-
F\left[ \frac{1}{k^2 \varepsilon_{\kappa=0}(k)+M^2}\right]_{z}
\right)
}
\notag \\
\label{eq11}
\end{eqnarray}

\noindent
where $F\left[ f(k) \right]_{z} = \int \frac{dk}{(2\pi)}
e^{i k z} f(k)$ is the Fourier transform with respect to the variable $z$
(see Appendix).

In the strong coupling limit $\alpha k \gg k^2$
and $k^2 \varepsilon_{\kappa=0}(k) = k \alpha$. 
In this case the string tension reads 

\begin{eqnarray*}
\sigma_s \simeq \frac{g^2 M^2}{\alpha^2}
\end{eqnarray*}

\noindent
It was shown in \cite{Yazawa} that the absence of a matter field leads to a
string tension $\sigma = M g^2 / 4\pi$.
One sees that a finite matter field coupled to the electromagnetic field
through $\alpha$ affects the string tension and screens the coupling 
between test particles \cite{NazarioSantiago}.
The Chern-Simons term no longer affects the confining
mass $M$ in this treatment of the non-frustrated Heisenberg model. 
Each spinon is tied to a flux 
proportional to $1/\kappa$ \cite{Dunne}. However the instanton flux is 
independent of the symmetries underlying the Heisenberg model, in other words 
it is not controlled by $\kappa$. 
The topological charges of instantons are not altered by the spinon flux 
$1/\kappa$. This leads to a string tension unaffected by the CS flux 
but screened by matter field.
The spinons remain confined and lead to the absence of 
paramagnetic phase in the non-frustrated Heisenberg model at zero temperature, 
with respect to this treatment.

\section{Conclusion}

We mapped a two dimensional Heisenberg Hamiltonian on an (2+1)-dimensional 
compact quantum electrodynamic Lagrangian with a Maxwell-Chern-Simons term at 
zero temperature. 
Here the spin site-occupation constraint is rigorously fixed by means of an 
imaginary chemical potential term\cite{Popov-88,PRB024409} which avoids  
the use a Lagrange multiplier constraint.

By symmetry consideration on a loop operator formed with the fermion operator
describing the spin arround a plaquette it is shown how a Chern-Simons
term enter the $QED_3$ description of the Heisenberg interaction.
The flux through the plaquette is fixed to multiples of $\pi$ in order to
enforce $SU(2)$ symmetry on the Heisenberg interaction.
The Chern-Simons action is introduced after taking the compact version
of the Maxwell-Spinon action in order to control the flux through
plaquettes formed by the spins.

We worked out the string tension of the confining potential which acts 
between the spinons and showed that the CS term induces a screening 
effect on the magnetic field. The confining potential between spinons
is affected by the matter field alone.

In conclusion we addressed the question about the possibility of 
controlling the deconfinement of spinons through flux affixed to them
and proportional to the inverse of the Chern-Simons parameter $\kappa$.
The confining string tension is not affected by the CS parameter even though
this is the case for the magnetic field. 
Our treatment agrees
with the fact that for an unfrustrated Heisenberg model the
spinon would not be in a deconfined phase \cite{NogueiraKleinert_condmat07}.
At zero temperature non-frustrated Heisenberg systems should not
present a paramagnetic phase.

A better treatment would be to take the matter-screen\-ed instanton flux 
$\phi_{inst}$ into account and fix to multiples of $\pi$ the total flux
through plaquette $\phi_\square$. This could possibly lead to consider
instanton configuration with winding charge greater than one as well
as a string tension depending on the spinon flux $1/\kappa$.

\section{Appendix}

\noindent
In the case of strong coupling $\alpha k \gg k^2$ the dielectric function
reads $k^2 \varepsilon_{\kappa}(k) = k \left(\alpha + \kappa^2/\alpha \right)$ 
and one gets

\begin{eqnarray*}
F\left[\frac{1}{k^2 \varepsilon_{\kappa=0}(k)} \right]_{z}
&=&
\frac{1}{\alpha}. i \theta(z)
 \\
F\left[\frac{1}{k^2 \varepsilon_{\kappa=0}(k) + M^2}\right]_{z}
&=&
\frac{1}{\alpha}. i \theta(z).
e^{-i\frac{M^2}{\alpha} z}
\end{eqnarray*}

\noindent
Here we choose to define the step function $\theta(z)$ as 

\begin{eqnarray*}
\theta(z) =
\begin{array}{|cc}
1 & z > 0 \\
0 & z \le 0
\end{array}
\end{eqnarray*}


\begin{thebibliography}{99}

\bibitem{LeeNagaosaWen-04} P.A. Lee, N. Nagaosa and X.-G. Wen,
Rev. Mod. Phys. 78, 17 (2006).

\bibitem{Senthil} T. Senthil, A. Vishwanath, L. Balents, S. Sachdev and
M.P.A. Fisher, Science \textbf{303}, 1490 (2004)

\bibitem{aff1} I. Affleck and J. B. Marston, Phys. Rev. {\bf B37}, 3774 (1988)

\bibitem{aff2} J. B. Marston and I. Affleck, Phys. Rev. {\bf B39}, 11538 (1989)

\bibitem{Marston} J. Brad Marston, Phys. Rev. Lett. \textbf{61}, 1914 (1988)

\bibitem{Popov-88} V. N. Popov and S. A. Fedotov, Sov. Phys. JETP 
\textbf{67}, 535 (1988)

\bibitem{PRB024409} 
R. Dillenschneider and J. Richert, Phys. Rev. B \textbf{73} 024409 (2006)

\bibitem{GhaemiSenthil-05} P. Ghaemi and T. Senthil, Phys. Rev. B \textbf{73}, 
054415 (2006)

\bibitem{PRB224443} 
R. Dillenschneider and J. Richert, Phys. Rev. B \textbf{73} 224443 (2006)

\bibitem{Morinari-05} T. Morinari, cond-mat/0508251 (2005)

\bibitem{Itzykson} C. Itsykson, J.-B. Zuber, Quantum Field Theory, McGraw-Hill,
(1986)

\bibitem{PRB144404}
R. Dillenschneider and J. Richert, Phys. Rev. B \textbf{74} 144404 (2006)

\bibitem{Polyakov} A. M. Polyakov, Gauge Fields and Strings, Harwood Academic
Publishers, (1987)

\bibitem{KleinertNogueira} F.S. Nogueira and H. Kleinert, Phys. Rev. Lett. 
\textbf{95} 176406 (2005)

\bibitem{NazarioSantiago} Z. Nazario and D.I. Santiago, cond-mat/0606386 (2006)

\bibitem{DeserJackiwTempleton} S. Deser, R. Jackiw and S. Templeton, Ann. Phys.
(NY) \textbf{140} 372 (1980).

\bibitem{Wilson} K.G. Wilson, Phys. Rev. D \textbf{10} 2445 (1974)

\bibitem{Nagaosav1} N. Nagaosa, Quantum field theory in condensed matter 
physics, Springer (1999)

\bibitem{Luscher} M. L\"{u}scher, Nucl. Phys. B \textbf{180}, 317 (1981)

\bibitem{Diamantini} M.C. Diamantini and P. Sodano, Phys. Rev. Lett. 
\textbf{71} 1969 (1993)

\bibitem{Yazawa} T. Yazawa and T. Suzuki, JHEP \textbf{04} 026 (2001)

\bibitem{Dunne} G.V. Dunne, Aspects of a Chern-Simons theory, hep-th/9902115
(1999)

\bibitem{NogueiraKleinert_condmat07} F.S. Nogueira and H. Kleinert, 
cond-mat/0705.3541 (2007).

\end{thebibliography}
\end{document}